# GENERAL EXPRESSIONS FOR CHERN FORMS UP TO THE $13^{th}$ ORDER IN CURVATURE

C. C. Briggs
*Center for Academic Computing, Penn State University, University Park, PA 16802*

**Abstract.** General expressions are given for Chern forms up to the $13^{th}$ order in curvature in terms of simple polynomial concomitants of the curvature 2-form for $n$-dimensional differentiable manifolds having a general linear connection.

This letter presents general expressions for Chern forms $c_{(p)}$ for $0 \le p \le 13$ in terms of simple polynomial concomitants of the curvature 2-form for $n$-dimensional differentiable manifolds having a general linear connection.

The $p^{th}$ Chern forms[1] $c_{(p)}$ representing the corresponding $p^{th}$ Chern classes of such a manifold $M$ can be defined by[2-3]

$$c_{(p)} \equiv \begin{cases} 1, & \text{if } p = 0 \\ \dfrac{i^p}{2^p \pi^p}\, \Omega_{[i_1}{}^{i_1} \wedge \Omega_{i_2}{}^{i_2} \wedge \ldots \wedge \Omega_{i_p]}{}^{i_p}, & \text{if } p > 0 \end{cases}, \qquad (1)$$

where $\Omega_a{}^b$ is the curvature 2-form of $M$.

Some numerical properties of $c_{(p)}$ for $0 \le p \le 13$ appear in Table 1. General expressions for $c_{(p)}$ for $0 \le p \le 13$ appear in Eqs. (3) through (16) in terms of the comcomitants $\mathrm{tr}^A(\Omega^B)$ defined by

$$\mathrm{tr}^A(\Omega^B) \equiv \Omega_{i_2}{}^{i_1} \wedge \Omega_{i_3}{}^{i_2} \wedge \ldots \wedge \Omega_{i_1}{}^{i_B} \wedge \Omega_{i_{B+2}}{}^{i_{B+1}} \wedge \Omega_{i_{B+3}}{}^{i_{B+2}} \wedge \ldots \wedge \Omega_{i_{B+1}}{}^{i_{2B}} \wedge \ldots \wedge \Omega_{i_{(A-1)B+2}}{}^{i_{(A-1)B+1}} \wedge \Omega_{i_{(A-1)B+3}}{}^{i_{(A-1)B+2}} \wedge \ldots \wedge \Omega_{i_{(A-1)B+1}}{}^{i_{AB}}, \qquad (2)$$

where $A$ and $B$ are integers $\ge 1$.

TABLE 1. SOME NUMERICAL PROPERTIES OF $c_{(p)}$ FOR $0 \le p \le 13$

| QUANTITY | ORDER | CURVATURE DEPENDENCE | MINIMUM NUMBER OF DIMENSIONS | NUMBER OF TERMS | $1^{st}$ OVERALL NUMERICAL FACTOR | $2^{nd}$ OVERALL NUMERICAL FACTOR | NUMBER OF PERMUTATIONS COMPREHENDED |
|---|---|---|---|---|---|---|---|
| $c_{(p)}$ | $p$ | — | $2p$ | — | $\frac{i^p}{2^p\pi^p}$ | $\frac{i^p}{2^p\pi^p p!}$ | $p!$ |
| $c_{(0)}$ | 0 | Zero | 0 | **1** | $+1$ | $+1$ | 1 |
| $c_{(1)}$ | 1 | Linear | 2 | **1** | $+\frac{1}{2\pi}i$ | $+\frac{1}{2\pi}i$ | 1 |
| $c_{(2)}$ | 2 | Quadratic | 4 | **2** | $-\frac{1}{4\pi^2}$ | $-\frac{1}{8\pi^2}$ | 2 |
| $c_{(3)}$ | 3 | Cubic | 6 | **3** | $-\frac{1}{8\pi^3}i$ | $-\frac{1}{48\pi^3}i$ | 6 |
| $c_{(4)}$ | 4 | Quartic | 8 | **5** | $+\frac{1}{16\pi^4}$ | $+\frac{1}{384\pi^4}$ | 24 |
| $c_{(5)}$ | 5 | Quintic | 10 | **7** | $+\frac{1}{32\pi^5}i$ | $+\frac{1}{3840\pi^5}i$ | 120 |
| $c_{(6)}$ | 6 | Sextic | 12 | **11** | $-\frac{1}{64\pi^6}$ | $-\frac{1}{46{,}080\pi^6}$ | 720 |
| $c_{(7)}$ | 7 | Septic | 14 | **15** | $-\frac{1}{128\pi^7}i$ | $-\frac{1}{645{,}120\pi^7}i$ | 5040 |
| $c_{(8)}$ | 8 | Octic | 16 | **22** | $+\frac{1}{256\pi^8}$ | $+\frac{1}{10{,}321{,}920\pi^8}$ | 40,320 |
| $c_{(9)}$ | 9 | Nonic | 18 | **30** | $+\frac{1}{512\pi^9}i$ | $+\frac{1}{185{,}794{,}560\pi^9}i$ | 362,880 |
| $c_{(10)}$ | 10 | Decic | 20 | **42** | $-\frac{1}{1024\pi^{10}}$ | $-\frac{1}{3{,}715{,}891{,}200\pi^{10}}$ | 3,628,800 |
| $c_{(11)}$ | 11 | Undecic | 22 | **56** | $-\frac{1}{2048\pi^{11}}i$ | $-\frac{1}{81{,}749{,}606{,}400\pi^{11}}i$ | 39,916,800 |
| $c_{(12)}$ | 12 | Duodecic | 24 | **77** | $+\frac{1}{4096\pi^{12}}$ | $+\frac{1}{1{,}961{,}990{,}553{,}600\pi^{12}}$ | 479,001,600 |
| $c_{(13)}$ | 13 | Tredecic | 26 | **101** | $+\frac{1}{8192\pi^{13}}i$ | $+\frac{1}{51{,}011{,}754{,}393{,}600\pi^{13}}i$ | 6,227,020,800 |

### $0^{th}$ CHERN FORM

$$c_{(0)} = \frac{i^0}{2^0\pi^0}(+1) = \frac{i^0}{2^0\pi^0}(+1) \qquad (3)$$

### $1^{st}$ CHERN FORM

$$c_{(1)} = \frac{i}{2\pi}\Omega_{i_1}{}^{i_1} = \frac{i}{2\pi}\mathrm{tr}(\Omega) \qquad (4)$$

### $2^{nd}$ CHERN FORM

$$c_{(2)} = \frac{i^2}{2^2\pi^2}\Omega_{[i_1}{}^{i_1} \wedge \Omega_{i_2]}{}^{i_2} = \frac{i^2}{2^2\pi^2 2!}(-\mathrm{tr}(\Omega^2) + \mathrm{tr}^2(\Omega)) \qquad (5)$$

### $3^{rd}$ CHERN FORM

$$c_{(3)} = \frac{i^3}{2^3\pi^3}\Omega_{[i_1}{}^{i_1} \wedge \Omega_{i_2}{}^{i_2} \wedge \Omega_{i_3]}{}^{i_3} = \frac{i^3}{2^3\pi^3 3!}(+2\,\mathrm{tr}(\Omega^3) - 3\,\mathrm{tr}(\Omega^2) \wedge \mathrm{tr}(\Omega) + \mathrm{tr}^3(\Omega)) \qquad (6)$$

### $4^{th}$ CHERN FORM

$$c_{(4)} = \frac{i^4}{2^4\pi^4}\Omega_{[i_1}{}^{i_1} \wedge \Omega_{i_2}{}^{i_2} \wedge \Omega_{i_3}{}^{i_3} \wedge \Omega_{i_4]}{}^{i_4} = \frac{i^4}{2^4\pi^4 4!}(-6\,\mathrm{tr}(\Omega^4) + 8\,\mathrm{tr}(\Omega^3) \wedge \mathrm{tr}(\Omega) + 3\,\mathrm{tr}^2(\Omega^2) - 6\,\mathrm{tr}(\Omega^2) \wedge \mathrm{tr}^2(\Omega) + \mathrm{tr}^4(\Omega)) \qquad (7)$$

### $5^{th}$ CHERN FORM

$$c_{(5)} = \frac{i^5}{2^5\pi^5}\Omega_{[i_1}{}^{i_1} \wedge \Omega_{i_2}{}^{i_2} \wedge \Omega_{i_3}{}^{i_3} \wedge \Omega_{i_4}{}^{i_4} \wedge \Omega_{i_5]}{}^{i_5} \qquad (8)$$

---

1 Eguchi, T., P. B. Gilkey, and A. J. Hanson, “Gravitation, gauge theories and differential geometry,” *Phys. Rep.*, **66** (1980) 213.
2 Kobayashi, S., and K. Nomizu, *Foundations of Differential Geometry*, John Wiley & Sons, New York (1969), p. 309.
3 Chern, S. S., “Characteristic classes of Hermitian manifolds,” *Annals of Math.*, **47** (1946) 85.

$$= \frac{i^5}{2^5\pi^5 5!}\big(+\,24\,\mathrm{tr}(\Omega^5) - 30\,\mathrm{tr}(\Omega^4)\wedge\mathrm{tr}(\Omega) - 20\,\mathrm{tr}(\Omega^3)\wedge\mathrm{tr}(\Omega^2) + 20\,\mathrm{tr}(\Omega^3)\wedge\mathrm{tr}^2(\Omega) + 15\,\mathrm{tr}^2(\Omega^2)\wedge\mathrm{tr}(\Omega) - 10\,\mathrm{tr}(\Omega^2)\wedge\mathrm{tr}^3(\Omega) + \mathrm{tr}^5(\Omega)\big)$$

### $6^{th}$ Chern Form

$$c_{(6)} = \frac{i^6}{2^6\pi^6}\,\Omega_{[i_1}{}^{i_1}\wedge\Omega_{i_2}{}^{i_2}\wedge\Omega_{i_3}{}^{i_3}\wedge\Omega_{i_4}{}^{i_4}\wedge\Omega_{i_5}{}^{i_5}\wedge\Omega_{i_6]}{}^{i_6} \tag{9}$$

$$= \frac{i^6}{2^6\pi^6 6!}\big(-\,120\,\mathrm{tr}(\Omega^6) + 144\,\mathrm{tr}(\Omega^5)\wedge\mathrm{tr}(\Omega) + 90\,\mathrm{tr}(\Omega^4)\wedge\mathrm{tr}(\Omega^2) - 90\,\mathrm{tr}(\Omega^4)\wedge\mathrm{tr}^2(\Omega) + 40\,\mathrm{tr}^2(\Omega^3) - 120\,\mathrm{tr}(\Omega^3)\wedge\mathrm{tr}(\Omega^2)\wedge\mathrm{tr}(\Omega) + 40\,\mathrm{tr}(\Omega^3)\wedge\mathrm{tr}^3(\Omega) - 15\,\mathrm{tr}^3(\Omega^2) + 45\,\mathrm{tr}^2(\Omega^2)\wedge\mathrm{tr}^2(\Omega) - 15\,\mathrm{tr}(\Omega^2)\wedge\mathrm{tr}^4(\Omega) + \mathrm{tr}^6(\Omega)\big)$$

### $7^{th}$ Chern Form

$$c_{(7)} = \frac{i^7}{2^7\pi^7}\,\Omega_{[i_1}{}^{i_1}\wedge\Omega_{i_2}{}^{i_2}\wedge\Omega_{i_3}{}^{i_3}\wedge\Omega_{i_4}{}^{i_4}\wedge\Omega_{i_5}{}^{i_5}\wedge\Omega_{i_6}{}^{i_6}\wedge\Omega_{i_7]}{}^{i_7} \tag{10}$$

$$= \frac{i^7}{2^7\pi^7 7!}\big(+\,720\,\mathrm{tr}(\Omega^7) - 840\,\mathrm{tr}(\Omega^6)\wedge\mathrm{tr}(\Omega) - 504\,\mathrm{tr}(\Omega^5)\wedge\mathrm{tr}(\Omega^2) + 504\,\mathrm{tr}(\Omega^5)\wedge\mathrm{tr}^2(\Omega) - 420\,\mathrm{tr}(\Omega^4)\wedge\mathrm{tr}(\Omega^3) + 630\,\mathrm{tr}(\Omega^4)\wedge\mathrm{tr}(\Omega^2)\wedge\mathrm{tr}(\Omega) - 210\,\mathrm{tr}(\Omega^4)\wedge\mathrm{tr}^3(\Omega) + 280\,\mathrm{tr}^2(\Omega^3)\wedge\mathrm{tr}(\Omega) + 210\,\mathrm{tr}(\Omega^3)\wedge\mathrm{tr}^2(\Omega^2) - 420\,\mathrm{tr}(\Omega^3)\wedge\mathrm{tr}(\Omega^2)\wedge\mathrm{tr}^2(\Omega) + 70\,\mathrm{tr}(\Omega^3)\wedge\mathrm{tr}^4(\Omega) - 105\,\mathrm{tr}^3(\Omega^2)\wedge\mathrm{tr}(\Omega) + 105\,\mathrm{tr}^2(\Omega^2)\wedge\mathrm{tr}^3(\Omega) - 21\,\mathrm{tr}(\Omega^2)\wedge\mathrm{tr}^5(\Omega) + \mathrm{tr}^7(\Omega)\big)$$

### $8^{th}$ Chern Form

$$c_{(8)} = \frac{i^8}{2^8\pi^8}\,\Omega_{[i_1}{}^{i_1}\wedge\Omega_{i_2}{}^{i_2}\wedge\Omega_{i_3}{}^{i_3}\wedge\Omega_{i_4}{}^{i_4}\wedge\Omega_{i_5}{}^{i_5}\wedge\Omega_{i_6}{}^{i_6}\wedge\Omega_{i_7}{}^{i_7}\wedge\Omega_{i_8]}{}^{i_8} \tag{11}$$

$$= \frac{i^8}{2^8\pi^8 8!}\big(-\,5040\,\mathrm{tr}(\Omega^8) + 5760\,\mathrm{tr}(\Omega^7)\wedge\mathrm{tr}(\Omega) + 3360\,\mathrm{tr}(\Omega^6)\wedge\mathrm{tr}(\Omega^2) - 3360\,\mathrm{tr}(\Omega^6)\wedge\mathrm{tr}^2(\Omega) + 2688\,\mathrm{tr}(\Omega^5)\wedge\mathrm{tr}(\Omega^3) - 4032\,\mathrm{tr}(\Omega^5)\wedge\mathrm{tr}(\Omega^2)\wedge\mathrm{tr}(\Omega) + 1344\,\mathrm{tr}(\Omega^5)\wedge\mathrm{tr}^3(\Omega) + 1260\,\mathrm{tr}^2(\Omega^4) - 3360\,\mathrm{tr}(\Omega^4)\wedge\mathrm{tr}(\Omega^3)\wedge\mathrm{tr}(\Omega) - 1260\,\mathrm{tr}(\Omega^4)\wedge\mathrm{tr}^2(\Omega^2) + 2520\,\mathrm{tr}(\Omega^4)\wedge\mathrm{tr}(\Omega^2)\wedge\mathrm{tr}^2(\Omega) - 420\,\mathrm{tr}(\Omega^4)\wedge\mathrm{tr}^4(\Omega) - 1120\,\mathrm{tr}^2(\Omega^3)\wedge\mathrm{tr}(\Omega^2) + 1120\,\mathrm{tr}^2(\Omega^3)\wedge\mathrm{tr}^2(\Omega) + 1680\,\mathrm{tr}(\Omega^3)\wedge\mathrm{tr}^2(\Omega^2)\wedge\mathrm{tr}(\Omega) - 1120\,\mathrm{tr}(\Omega^3)\wedge\mathrm{tr}(\Omega^2)\wedge\mathrm{tr}^3(\Omega) + 112\,\mathrm{tr}(\Omega^3)\wedge\mathrm{tr}^5(\Omega) + 105\,\mathrm{tr}^4(\Omega^2) - 420\,\mathrm{tr}^3(\Omega^2)\wedge\mathrm{tr}^2(\Omega) + 210\,\mathrm{tr}^2(\Omega^2)\wedge\mathrm{tr}^4(\Omega) - 28\,\mathrm{tr}(\Omega^2)\wedge\mathrm{tr}^6(\Omega) + \mathrm{tr}^8(\Omega)\big)$$

### $9^{th}$ Chern Form

$$c_{(9)} = \frac{i^9}{2^9\pi^9}\,\Omega_{[i_1}{}^{i_1}\wedge\Omega_{i_2}{}^{i_2}\wedge\Omega_{i_3}{}^{i_3}\wedge\Omega_{i_4}{}^{i_4}\wedge\Omega_{i_5}{}^{i_5}\wedge\Omega_{i_6}{}^{i_6}\wedge\Omega_{i_7}{}^{i_7}\wedge\Omega_{i_8}{}^{i_8}\wedge\Omega_{i_9]}{}^{i_9} \tag{12}$$

$$= \frac{i^9}{2^9\pi^9 9!}\big(+\,40{,}320\,\mathrm{tr}(\Omega^9) - 45{,}360\,\mathrm{tr}(\Omega^8)\wedge\mathrm{tr}(\Omega) - 25{,}920\,\mathrm{tr}(\Omega^7)\wedge\mathrm{tr}(\Omega^2) + 25{,}920\,\mathrm{tr}(\Omega^7)\wedge\mathrm{tr}^2(\Omega) - 20{,}160\,\mathrm{tr}(\Omega^6)\wedge\mathrm{tr}(\Omega^3) + 30{,}240\,\mathrm{tr}(\Omega^6)\wedge\mathrm{tr}(\Omega^2)\wedge\mathrm{tr}(\Omega) - 10{,}080\,\mathrm{tr}(\Omega^6)\wedge\mathrm{tr}^3(\Omega) - 18{,}144\,\mathrm{tr}(\Omega^5)\wedge\mathrm{tr}(\Omega^4) + 24{,}192\,\mathrm{tr}(\Omega^5)\wedge\mathrm{tr}(\Omega^3)\wedge\mathrm{tr}(\Omega) + 9072\,\mathrm{tr}(\Omega^5)\wedge\mathrm{tr}^2(\Omega^2) - 18{,}144\,\mathrm{tr}(\Omega^5)\wedge\mathrm{tr}(\Omega^2)\wedge\mathrm{tr}^2(\Omega) + 3024\,\mathrm{tr}(\Omega^5)\wedge\mathrm{tr}^4(\Omega) + 11{,}340\,\mathrm{tr}^2(\Omega^4)\wedge\mathrm{tr}(\Omega) + 15{,}120\,\mathrm{tr}(\Omega^4)\wedge\mathrm{tr}(\Omega^3)\wedge\mathrm{tr}(\Omega^2) - 15{,}120\,\mathrm{tr}(\Omega^4)\wedge\mathrm{tr}(\Omega^3)\wedge\mathrm{tr}^2(\Omega) - 11{,}340\,\mathrm{tr}(\Omega^4)\wedge\mathrm{tr}^2(\Omega^2)\wedge\mathrm{tr}(\Omega) + 7560\,\mathrm{tr}(\Omega^4)\wedge\mathrm{tr}(\Omega^2)\wedge\mathrm{tr}^3(\Omega) - 756\,\mathrm{tr}(\Omega^4)\wedge\mathrm{tr}^5(\Omega) + 2240\,\mathrm{tr}^3(\Omega^3) - 10{,}080\,\mathrm{tr}^2(\Omega^3)\wedge\mathrm{tr}(\Omega^2)\wedge\mathrm{tr}(\Omega) + 3360\,\mathrm{tr}^2(\Omega^3)\wedge\mathrm{tr}^3(\Omega) - 2520\,\mathrm{tr}(\Omega^3)\wedge\mathrm{tr}^3(\Omega^2) + 7560\,\mathrm{tr}(\Omega^3)\wedge\mathrm{tr}^2(\Omega^2)\wedge\mathrm{tr}^2(\Omega) - 2520\,\mathrm{tr}(\Omega^3)\wedge\mathrm{tr}(\Omega^2)\wedge\mathrm{tr}^4(\Omega) + 168\,\mathrm{tr}(\Omega^3)\wedge\mathrm{tr}^6(\Omega) + 945\,\mathrm{tr}^4(\Omega^2)\wedge\mathrm{tr}(\Omega) - 1260\,\mathrm{tr}^3(\Omega^2)\wedge\mathrm{tr}^3(\Omega) + 378\,\mathrm{tr}^2(\Omega^2)\wedge\mathrm{tr}^5(\Omega) - 36\,\mathrm{tr}(\Omega^2)\wedge\mathrm{tr}^7(\Omega) + \mathrm{tr}^9(\Omega)\big)$$

### $10^{th}$ Chern Form

$$c_{(10)} = \frac{i^{10}}{2^{10}\pi^{10}}\,\Omega_{[i_1}{}^{i_1}\wedge\Omega_{i_2}{}^{i_2}\wedge\Omega_{i_3}{}^{i_3}\wedge\Omega_{i_4}{}^{i_4}\wedge\Omega_{i_5}{}^{i_5}\wedge\Omega_{i_6}{}^{i_6}\wedge\Omega_{i_7}{}^{i_7}\wedge\Omega_{i_8}{}^{i_8}\wedge\Omega_{i_9}{}^{i_9}\wedge\Omega_{i_{10}]}{}^{i_{10}} \tag{13}$$

$$= \frac{i^{10}}{2^{10}\pi^{10} 10!}\big(-\,362{,}880\,\mathrm{tr}(\Omega^{10}) + 403{,}200\,\mathrm{tr}(\Omega^9)\wedge\mathrm{tr}(\Omega) + 226{,}800\,\mathrm{tr}(\Omega^8)\wedge\mathrm{tr}(\Omega^2) - 226{,}800\,\mathrm{tr}(\Omega^8)\wedge\mathrm{tr}^2(\Omega) + 172{,}800\,\mathrm{tr}(\Omega^7)\wedge\mathrm{tr}(\Omega^3) - 259{,}200\,\mathrm{tr}(\Omega^7)\wedge\mathrm{tr}(\Omega^2)\wedge\mathrm{tr}(\Omega) + 86{,}400\,\mathrm{tr}(\Omega^7)\wedge\mathrm{tr}^3(\Omega) + 151200\,\mathrm{tr}(\Omega^6)\wedge\mathrm{tr}(\Omega^4) - 201{,}600\,\mathrm{tr}(\Omega^6)\wedge\mathrm{tr}(\Omega^3)\wedge\mathrm{tr}(\Omega) - 75{,}600\,\mathrm{tr}(\Omega^6)\wedge\mathrm{tr}^2(\Omega^2) + 151{,}200\,\mathrm{tr}(\Omega^6)\wedge\mathrm{tr}(\Omega^2)\wedge\mathrm{tr}^2(\Omega) - 25{,}200\,\mathrm{tr}(\Omega^6)\wedge\mathrm{tr}^4(\Omega) + 72{,}576\,\mathrm{tr}^2(\Omega^5) - 181{,}440\,\mathrm{tr}(\Omega^5)\wedge\mathrm{tr}(\Omega^4)\wedge\mathrm{tr}(\Omega) - 120{,}960\,\mathrm{tr}(\Omega^5)\wedge\mathrm{tr}(\Omega^3)\wedge\mathrm{tr}(\Omega^2) + 120{,}960\,\mathrm{tr}(\Omega^5)\wedge\mathrm{tr}(\Omega^3)\wedge\mathrm{tr}^2(\Omega) + 90{,}720\,\mathrm{tr}(\Omega^5)\wedge\mathrm{tr}^2(\Omega^2)\wedge\mathrm{tr}(\Omega) - 60{,}480\,\mathrm{tr}(\Omega^5)\wedge\mathrm{tr}(\Omega^2)\wedge\mathrm{tr}^3(\Omega) + 6048\,\mathrm{tr}(\Omega^5)\wedge\mathrm{tr}^5(\Omega) - 56{,}700\,\mathrm{tr}^2(\Omega^4)\wedge\mathrm{tr}(\Omega^2) + 56{,}700\,\mathrm{tr}^2(\Omega^4)\wedge\mathrm{tr}^2(\Omega) - 50{,}400\,\mathrm{tr}(\Omega^4)\wedge\mathrm{tr}^2(\Omega^3) + 151{,}200\,\mathrm{tr}(\Omega^4)\wedge\mathrm{tr}(\Omega^3)\wedge\mathrm{tr}(\Omega^2)\wedge\mathrm{tr}(\Omega) - 50{,}400\,\mathrm{tr}(\Omega^4)\wedge\mathrm{tr}(\Omega^3)\wedge\mathrm{tr}^3(\Omega) + 18{,}900\,\mathrm{tr}(\Omega^4)\wedge\mathrm{tr}^3(\Omega^2) - 56{,}700\,\mathrm{tr}(\Omega^4)\wedge\mathrm{tr}^2(\Omega^2)\wedge\mathrm{tr}^2(\Omega) + 18{,}900\,\mathrm{tr}(\Omega^4)\wedge\mathrm{tr}(\Omega^2)\wedge\mathrm{tr}^4(\Omega) - 1260\,\mathrm{tr}(\Omega^4)\wedge\mathrm{tr}^6(\Omega) + 22{,}400\,\mathrm{tr}^3(\Omega^3)\wedge\mathrm{tr}(\Omega) + 25{,}200\,\mathrm{tr}^2(\Omega^3)\wedge\mathrm{tr}^2(\Omega^2) - 50{,}400\,\mathrm{tr}^2(\Omega^3)\wedge\mathrm{tr}(\Omega^2)\wedge\mathrm{tr}^2(\Omega) + 8400\,\mathrm{tr}^2(\Omega^3)\wedge\mathrm{tr}^4(\Omega) - 25{,}200\,\mathrm{tr}(\Omega^3)\wedge\mathrm{tr}^3(\Omega^2)\wedge\mathrm{tr}(\Omega) + 25{,}200\,\mathrm{tr}(\Omega^3)\wedge\mathrm{tr}^2(\Omega^2)\wedge\mathrm{tr}^3(\Omega) - 5040\,\mathrm{tr}(\Omega^3)\wedge\mathrm{tr}(\Omega^2)\wedge\mathrm{tr}^5(\Omega) + 240\,\mathrm{tr}(\Omega^3)\wedge\mathrm{tr}^7(\Omega) - 945\,\mathrm{tr}^5(\Omega^2) + 4725\,\mathrm{tr}^4(\Omega^2)\wedge\mathrm{tr}^2(\Omega) - 3150\,\mathrm{tr}^3(\Omega^2)\wedge\mathrm{tr}^4(\Omega) + 630\,\mathrm{tr}^2(\Omega^2)\wedge\mathrm{tr}^6(\Omega) - 45\,\mathrm{tr}(\Omega^2)\wedge\mathrm{tr}^8(\Omega) + \mathrm{tr}^{10}(\Omega)\big)$$

## $11^{th}$ Chern Form

$$c_{(11)} = \frac{i^{11}}{2^{11}\pi^{11}}\, \Omega_{[i_1}{}^{i_1} \wedge \Omega_{i_2}{}^{i_2} \wedge \Omega_{i_3}{}^{i_3} \wedge \Omega_{i_4}{}^{i_4} \wedge \Omega_{i_5}{}^{i_5} \wedge \Omega_{i_6}{}^{i_6} \wedge \Omega_{i_7}{}^{i_7} \wedge \Omega_{i_8}{}^{i_8} \wedge \Omega_{i_9}{}^{i_9} \wedge \Omega_{i_{10}}{}^{i_{10}} \wedge \Omega_{i_{11}]}{}^{i_{11}} \tag{14}$$

$$= \frac{i^{11}}{2^{11}\pi^{11}11!} (+\, 3{,}628{,}800\, \mathrm{tr}(\Omega^{11}) - 3{,}991{,}680\, \mathrm{tr}(\Omega^{10}) \wedge \mathrm{tr}(\Omega) - 2{,}217{,}600\, \mathrm{tr}(\Omega^{9}) \wedge \mathrm{tr}(\Omega^{2}) + 2{,}217{,}600\, \mathrm{tr}(\Omega^{9}) \wedge \mathrm{tr}^2(\Omega) -$$

$$- 1{,}663{,}200\, \mathrm{tr}(\Omega^{8}) \wedge \mathrm{tr}(\Omega^{3}) + 2{,}494{,}800\, \mathrm{tr}(\Omega^{8}) \wedge \mathrm{tr}(\Omega^{2}) \wedge \mathrm{tr}(\Omega) - 831{,}600\, \mathrm{tr}(\Omega^{8}) \wedge \mathrm{tr}^3(\Omega) - 1{,}425{,}600\, \mathrm{tr}(\Omega^{7}) \wedge \mathrm{tr}(\Omega^{4}) +$$

$$+ 1{,}900{,}800\, \mathrm{tr}(\Omega^{7}) \wedge \mathrm{tr}(\Omega^{3}) \wedge \mathrm{tr}(\Omega) + 712{,}800\, \mathrm{tr}(\Omega^{7}) \wedge \mathrm{tr}^2(\Omega^{2}) - 1{,}425{,}600\, \mathrm{tr}(\Omega^{7}) \wedge \mathrm{tr}(\Omega^{2}) \wedge \mathrm{tr}^2(\Omega) + 237{,}600\, \mathrm{tr}(\Omega^{7}) \wedge \mathrm{tr}^4(\Omega) -$$

$$- 1{,}330{,}560\, \mathrm{tr}(\Omega^{6}) \wedge \mathrm{tr}(\Omega^{5}) + 1{,}663{,}200\, \mathrm{tr}(\Omega^{6}) \wedge \mathrm{tr}(\Omega^{4}) \wedge \mathrm{tr}(\Omega) + 1{,}108{,}800\, \mathrm{tr}(\Omega^{6}) \wedge \mathrm{tr}(\Omega^{3}) \wedge \mathrm{tr}(\Omega^{2}) -$$

$$- 1{,}108{,}800\, \mathrm{tr}(\Omega^{6}) \wedge \mathrm{tr}(\Omega^{3}) \wedge \mathrm{tr}^2(\Omega) - 831{,}600\, \mathrm{tr}(\Omega^{6}) \wedge \mathrm{tr}^2(\Omega^{2}) \wedge \mathrm{tr}(\Omega) + 554{,}400\, \mathrm{tr}(\Omega^{6}) \wedge \mathrm{tr}(\Omega^{2}) \wedge \mathrm{tr}^3(\Omega) -$$

$$- 55{,}440\, \mathrm{tr}(\Omega^{6}) \wedge \mathrm{tr}^5(\Omega) + 798{,}336\, \mathrm{tr}^2(\Omega^{5}) \wedge \mathrm{tr}(\Omega) + 997{,}920\, \mathrm{tr}(\Omega^{5}) \wedge \mathrm{tr}(\Omega^{4}) \wedge \mathrm{tr}(\Omega^{2}) - 997{,}920\, \mathrm{tr}(\Omega^{5}) \wedge \mathrm{tr}(\Omega^{4}) \wedge \mathrm{tr}^2(\Omega) +$$

$$+ 443{,}520\, \mathrm{tr}(\Omega^{5}) \wedge \mathrm{tr}^2(\Omega^{3}) - 1{,}330{,}560\, \mathrm{tr}(\Omega^{5}) \wedge \mathrm{tr}(\Omega^{3}) \wedge \mathrm{tr}(\Omega^{2}) \wedge \mathrm{tr}(\Omega) + 443{,}520\, \mathrm{tr}(\Omega^{5}) \wedge \mathrm{tr}(\Omega^{3}) \wedge \mathrm{tr}^3(\Omega) -$$

$$- 166{,}320\, \mathrm{tr}(\Omega^{5}) \wedge \mathrm{tr}^3(\Omega^{2}) + 498{,}960\, \mathrm{tr}(\Omega^{5}) \wedge \mathrm{tr}^2(\Omega^{2}) \wedge \mathrm{tr}^2(\Omega) - 166{,}320\, \mathrm{tr}(\Omega^{5}) \wedge \mathrm{tr}(\Omega^{2}) \wedge \mathrm{tr}^4(\Omega) + 11{,}088\, \mathrm{tr}(\Omega^{5}) \wedge \mathrm{tr}^6(\Omega) +$$

$$+ 415{,}800\, \mathrm{tr}^2(\Omega^{4}) \wedge \mathrm{tr}(\Omega^{3}) - 623{,}700\, \mathrm{tr}^2(\Omega^{4}) \wedge \mathrm{tr}(\Omega^{2}) \wedge \mathrm{tr}(\Omega) + 207{,}900\, \mathrm{tr}^2(\Omega^{4}) \wedge \mathrm{tr}^3(\Omega) - 554{,}400\, \mathrm{tr}(\Omega^{4}) \wedge \mathrm{tr}^2(\Omega^{3}) \wedge \mathrm{tr}(\Omega) -$$

$$- 415{,}800\, \mathrm{tr}(\Omega^{4}) \wedge \mathrm{tr}(\Omega^{3}) \wedge \mathrm{tr}^2(\Omega^{2}) + 831{,}600\, \mathrm{tr}(\Omega^{4}) \wedge \mathrm{tr}(\Omega^{3}) \wedge \mathrm{tr}(\Omega^{2}) \wedge \mathrm{tr}^2(\Omega) - 138{,}600\, \mathrm{tr}(\Omega^{4}) \wedge \mathrm{tr}(\Omega^{3}) \wedge \mathrm{tr}^4(\Omega) +$$

$$+ 207{,}900\, \mathrm{tr}(\Omega^{4}) \wedge \mathrm{tr}^3(\Omega^{2}) \wedge \mathrm{tr}(\Omega) - 207{,}900\, \mathrm{tr}(\Omega^{4}) \wedge \mathrm{tr}^2(\Omega^{2}) \wedge \mathrm{tr}^3(\Omega) + 41{,}580\, \mathrm{tr}(\Omega^{4}) \wedge \mathrm{tr}(\Omega^{2}) \wedge \mathrm{tr}^5(\Omega) - 1980\, \mathrm{tr}(\Omega^{4}) \wedge \mathrm{tr}^7(\Omega) -$$

$$- 123{,}200\, \mathrm{tr}^3(\Omega^{3}) \wedge \mathrm{tr}(\Omega^{2}) + 123{,}200\, \mathrm{tr}^3(\Omega^{3}) \wedge \mathrm{tr}^2(\Omega) + 277{,}200\, \mathrm{tr}^2(\Omega^{3}) \wedge \mathrm{tr}^2(\Omega^{2}) \wedge \mathrm{tr}(\Omega) - 184{,}800\, \mathrm{tr}^2(\Omega^{3}) \wedge \mathrm{tr}(\Omega^{2}) \wedge \mathrm{tr}^3(\Omega) +$$

$$+ 18{,}480\, \mathrm{tr}^2(\Omega^{3}) \wedge \mathrm{tr}^5(\Omega) + 34{,}650\, \mathrm{tr}(\Omega^{3}) \wedge \mathrm{tr}^4(\Omega^{2}) - 138{,}600\, \mathrm{tr}(\Omega^{3}) \wedge \mathrm{tr}^3(\Omega^{2}) \wedge \mathrm{tr}^2(\Omega) + 69{,}300\, \mathrm{tr}(\Omega^{3}) \wedge \mathrm{tr}^2(\Omega^{2}) \wedge \mathrm{tr}^4(\Omega) -$$

$$- 9240\, \mathrm{tr}(\Omega^{3}) \wedge \mathrm{tr}(\Omega^{2}) \wedge \mathrm{tr}^6(\Omega) + 330\, \mathrm{tr}(\Omega^{3}) \wedge \mathrm{tr}^8(\Omega) - 10{,}395\, \mathrm{tr}^5(\Omega^{2}) \wedge \mathrm{tr}(\Omega) + 17{,}325\, \mathrm{tr}^4(\Omega^{2}) \wedge \mathrm{tr}^3(\Omega) -$$

$$- 6930\, \mathrm{tr}^3(\Omega^{2}) \wedge \mathrm{tr}^5(\Omega) + 990\, \mathrm{tr}^2(\Omega^{2}) \wedge \mathrm{tr}^7(\Omega) - 55\, \mathrm{tr}(\Omega^{2}) \wedge \mathrm{tr}^9(\Omega) + \mathrm{tr}^{11}(\Omega))$$

## $12^{th}$ Chern Form

$$c_{(12)} = \frac{i^{12}}{2^{12}\pi^{12}}\, \Omega_{[i_1}{}^{i_1} \wedge \Omega_{i_2}{}^{i_2} \wedge \Omega_{i_3}{}^{i_3} \wedge \Omega_{i_4}{}^{i_4} \wedge \Omega_{i_5}{}^{i_5} \wedge \Omega_{i_6}{}^{i_6} \wedge \Omega_{i_7}{}^{i_7} \wedge \Omega_{i_8}{}^{i_8} \wedge \Omega_{i_9}{}^{i_9} \wedge \Omega_{i_{10}}{}^{i_{10}} \wedge \Omega_{i_{11}}{}^{i_{11}} \wedge \Omega_{i_{12}]}{}^{i_{12}} \tag{15}$$

$$= \frac{i^{12}}{2^{12}\pi^{12}12!} (-\, 39{,}916{,}800\, \mathrm{tr}(\Omega^{12}) + 43{,}545{,}600\, \mathrm{tr}(\Omega^{11}) \wedge \mathrm{tr}(\Omega) + 23{,}950{,}080\, \mathrm{tr}(\Omega^{10}) \wedge \mathrm{tr}(\Omega^{2}) - 23{,}950{,}080\, \mathrm{tr}(\Omega^{10}) \wedge \mathrm{tr}^2(\Omega) +$$

$$+ 17{,}740{,}800\, \mathrm{tr}(\Omega^{9}) \wedge \mathrm{tr}(\Omega^{3}) - 26{,}611{,}200\, \mathrm{tr}(\Omega^{9}) \wedge \mathrm{tr}(\Omega^{2}) \wedge \mathrm{tr}(\Omega) + 8{,}870{,}400\, \mathrm{tr}(\Omega^{9}) \wedge \mathrm{tr}^3(\Omega) + 14{,}968{,}800\, \mathrm{tr}(\Omega^{8}) \wedge \mathrm{tr}(\Omega^{4}) -$$

$$- 19{,}958{,}400\, \mathrm{tr}(\Omega^{8}) \wedge \mathrm{tr}(\Omega^{3}) \wedge \mathrm{tr}(\Omega) - 7{,}484{,}400\, \mathrm{tr}(\Omega^{8}) \wedge \mathrm{tr}^2(\Omega^{2}) + 14{,}968{,}800\, \mathrm{tr}(\Omega^{8}) \wedge \mathrm{tr}(\Omega^{2}) \wedge \mathrm{tr}^2(\Omega) -$$

$$- 2{,}494{,}800\, \mathrm{tr}(\Omega^{8}) \wedge \mathrm{tr}^4(\Omega) + 13{,}685{,}760\, \mathrm{tr}(\Omega^{7}) \wedge \mathrm{tr}(\Omega^{5}) - 17{,}107{,}200\, \mathrm{tr}(\Omega^{7}) \wedge \mathrm{tr}(\Omega^{4}) \wedge \mathrm{tr}(\Omega) -$$

$$- 11{,}404{,}800\, \mathrm{tr}(\Omega^{7}) \wedge \mathrm{tr}(\Omega^{3}) \wedge \mathrm{tr}(\Omega^{2}) + 11{,}404{,}800\, \mathrm{tr}(\Omega^{7}) \wedge \mathrm{tr}(\Omega^{3}) \wedge \mathrm{tr}^2(\Omega) + 8{,}553{,}600\, \mathrm{tr}(\Omega^{7}) \wedge \mathrm{tr}^2(\Omega^{2}) \wedge \mathrm{tr}(\Omega) -$$

$$- 5{,}702{,}400\, \mathrm{tr}(\Omega^{7}) \wedge \mathrm{tr}(\Omega^{2}) \wedge \mathrm{tr}^3(\Omega) + 570{,}240\, \mathrm{tr}(\Omega^{7}) \wedge \mathrm{tr}^5(\Omega) + 6{,}652{,}800\, \mathrm{tr}^2(\Omega^{6}) - 15{,}966{,}720\, \mathrm{tr}(\Omega^{6}) \wedge \mathrm{tr}(\Omega^{5}) \wedge \mathrm{tr}(\Omega) -$$

$$- 9{,}979{,}200\, \mathrm{tr}(\Omega^{6}) \wedge \mathrm{tr}(\Omega^{4}) \wedge \mathrm{tr}(\Omega^{2}) + 9{,}979{,}200\, \mathrm{tr}(\Omega^{6}) \wedge \mathrm{tr}(\Omega^{4}) \wedge \mathrm{tr}^2(\Omega) - 4{,}435{,}200\, \mathrm{tr}(\Omega^{6}) \wedge \mathrm{tr}^2(\Omega^{3}) +$$

$$+ 13{,}305{,}600\, \mathrm{tr}(\Omega^{6}) \wedge \mathrm{tr}(\Omega^{3}) \wedge \mathrm{tr}(\Omega^{2}) \wedge \mathrm{tr}(\Omega) - 4{,}435{,}200\, \mathrm{tr}(\Omega^{6}) \wedge \mathrm{tr}(\Omega^{3}) \wedge \mathrm{tr}^3(\Omega) + 1{,}663{,}200\, \mathrm{tr}(\Omega^{6}) \wedge \mathrm{tr}^3(\Omega^{2}) -$$

$$- 4{,}989{,}600\, \mathrm{tr}(\Omega^{6}) \wedge \mathrm{tr}^2(\Omega^{2}) \wedge \mathrm{tr}^2(\Omega) + 1{,}663{,}200\, \mathrm{tr}(\Omega^{6}) \wedge \mathrm{tr}(\Omega^{2}) \wedge \mathrm{tr}^4(\Omega) - 110{,}880\, \mathrm{tr}(\Omega^{6}) \wedge \mathrm{tr}^6(\Omega) -$$

$$- 4{,}790{,}016\, \mathrm{tr}^2(\Omega^{5}) \wedge \mathrm{tr}(\Omega^{2}) + 4{,}790{,}016\, \mathrm{tr}^2(\Omega^{5}) \wedge \mathrm{tr}^2(\Omega) - 7{,}983{,}360\, \mathrm{tr}(\Omega^{5}) \wedge \mathrm{tr}(\Omega^{4}) \wedge \mathrm{tr}(\Omega^{3}) +$$

$$+ 11{,}975{,}040\, \mathrm{tr}(\Omega^{5}) \wedge \mathrm{tr}(\Omega^{4}) \wedge \mathrm{tr}(\Omega^{2}) \wedge \mathrm{tr}(\Omega) - 3{,}991{,}680\, \mathrm{tr}(\Omega^{5}) \wedge \mathrm{tr}(\Omega^{4}) \wedge \mathrm{tr}^3(\Omega) + 5{,}322{,}240\, \mathrm{tr}(\Omega^{5}) \wedge \mathrm{tr}^2(\Omega^{3}) \wedge \mathrm{tr}(\Omega) +$$

$$+ 3{,}991{,}680\, \mathrm{tr}(\Omega^{5}) \wedge \mathrm{tr}(\Omega^{3}) \wedge \mathrm{tr}^2(\Omega^{2}) - 7{,}983{,}360\, \mathrm{tr}(\Omega^{5}) \wedge \mathrm{tr}(\Omega^{3}) \wedge \mathrm{tr}(\Omega^{2}) \wedge \mathrm{tr}^2(\Omega) + 1{,}330{,}560\, \mathrm{tr}(\Omega^{5}) \wedge \mathrm{tr}(\Omega^{3}) \wedge \mathrm{tr}^4(\Omega) -$$

$$- 1{,}995{,}840\, \mathrm{tr}(\Omega^{5}) \wedge \mathrm{tr}^3(\Omega^{2}) \wedge \mathrm{tr}(\Omega) + 1{,}995{,}840\, \mathrm{tr}(\Omega^{5}) \wedge \mathrm{tr}^2(\Omega^{2}) \wedge \mathrm{tr}^3(\Omega) - 399{,}168\, \mathrm{tr}(\Omega^{5}) \wedge \mathrm{tr}(\Omega^{2}) \wedge \mathrm{tr}^5(\Omega) +$$

$$+ 19{,}008\, \mathrm{tr}(\Omega^{5}) \wedge \mathrm{tr}^7(\Omega) - 1{,}247{,}400\, \mathrm{tr}^3(\Omega^{4}) + 4{,}989{,}600\, \mathrm{tr}^2(\Omega^{4}) \wedge \mathrm{tr}(\Omega^{3}) \wedge \mathrm{tr}(\Omega) + 1{,}871{,}100\, \mathrm{tr}^2(\Omega^{4}) \wedge \mathrm{tr}^2(\Omega^{2}) -$$

$$- 3{,}742{,}200\, \mathrm{tr}^2(\Omega^{4}) \wedge \mathrm{tr}(\Omega^{2}) \wedge \mathrm{tr}^2(\Omega) + 623{,}700\, \mathrm{tr}^2(\Omega^{4}) \wedge \mathrm{tr}^4(\Omega) + 3{,}326{,}400\, \mathrm{tr}(\Omega^{4}) \wedge \mathrm{tr}^2(\Omega^{3}) \wedge \mathrm{tr}(\Omega^{2}) -$$

$$- 3{,}326{,}400\, \mathrm{tr}(\Omega^{4}) \wedge \mathrm{tr}^2(\Omega^{3}) \wedge \mathrm{tr}^2(\Omega) - 4{,}989{,}600\, \mathrm{tr}(\Omega^{4}) \wedge \mathrm{tr}(\Omega^{3}) \wedge \mathrm{tr}^2(\Omega^{2}) \wedge \mathrm{tr}(\Omega) + 3{,}326{,}400\, \mathrm{tr}(\Omega^{4}) \wedge \mathrm{tr}(\Omega^{3}) \wedge \mathrm{tr}(\Omega^{2}) \wedge \mathrm{tr}^3(\Omega) -$$

$$- 332{,}640\, \mathrm{tr}(\Omega^{4}) \wedge \mathrm{tr}(\Omega^{3}) \wedge \mathrm{tr}^5(\Omega) - 311{,}850\, \mathrm{tr}(\Omega^{4}) \wedge \mathrm{tr}^4(\Omega^{2}) + 1{,}247{,}400\, \mathrm{tr}(\Omega^{4}) \wedge \mathrm{tr}^3(\Omega^{2}) \wedge \mathrm{tr}^2(\Omega) -$$

$$- 623{,}700\, \mathrm{tr}(\Omega^{4}) \wedge \mathrm{tr}^2(\Omega^{2}) \wedge \mathrm{tr}^4(\Omega) + 83{,}160\, \mathrm{tr}(\Omega^{4}) \wedge \mathrm{tr}(\Omega^{2}) \wedge \mathrm{tr}^6(\Omega) - 2970\, \mathrm{tr}(\Omega^{4}) \wedge \mathrm{tr}^8(\Omega) + 246{,}400\, \mathrm{tr}^4(\Omega^{3}) -$$

$$- 1{,}478{,}400\, \mathrm{tr}^3(\Omega^{3}) \wedge \mathrm{tr}(\Omega^{2}) \wedge \mathrm{tr}(\Omega) + 492{,}800\, \mathrm{tr}^3(\Omega^{3}) \wedge \mathrm{tr}^3(\Omega) - 554{,}400\, \mathrm{tr}^2(\Omega^{3}) \wedge \mathrm{tr}^3(\Omega^{2}) + 1{,}663{,}200\, \mathrm{tr}^2(\Omega^{3}) \wedge \mathrm{tr}^2(\Omega^{2}) \wedge \mathrm{tr}^2(\Omega) -$$

$$- 554{,}400\, \mathrm{tr}^2(\Omega^{3}) \wedge \mathrm{tr}(\Omega^{2}) \wedge \mathrm{tr}^4(\Omega) + 36{,}960\, \mathrm{tr}^2(\Omega^{3}) \wedge \mathrm{tr}^6(\Omega) + 415{,}800\, \mathrm{tr}(\Omega^{3}) \wedge \mathrm{tr}^4(\Omega^{2}) \wedge \mathrm{tr}(\Omega) - 554{,}400\, \mathrm{tr}(\Omega^{3}) \wedge \mathrm{tr}^3(\Omega^{2}) \wedge \mathrm{tr}^3(\Omega) +$$

$$+ 166{,}320\, \mathrm{tr}(\Omega^{3}) \wedge \mathrm{tr}^2(\Omega^{2}) \wedge \mathrm{tr}^5(\Omega) - 15{,}840\, \mathrm{tr}(\Omega^{3}) \wedge \mathrm{tr}(\Omega^{2}) \wedge \mathrm{tr}^7(\Omega) + 440\, \mathrm{tr}(\Omega^{3}) \wedge \mathrm{tr}^9(\Omega) + 10{,}395\, \mathrm{tr}^6(\Omega^{2}) - 62{,}370\, \mathrm{tr}^5(\Omega^{2}) \wedge \mathrm{tr}^2(\Omega) +$$

$$+\,51{,}975\ \mathrm{tr}^4(\Omega^2)\wedge\mathrm{tr}^4(\Omega) - 13{,}860\ \mathrm{tr}^3(\Omega^2)\wedge\mathrm{tr}^6(\Omega) + 1485\ \mathrm{tr}^2(\Omega^2)\wedge\mathrm{tr}^8(\Omega) - 66\ \mathrm{tr}(\Omega^2)\wedge\mathrm{tr}^{10}(\Omega) + \mathrm{tr}^{12}(\Omega))$$

## $13^{th}$ Chern Form

$$\begin{aligned}
c_{(13)} &= \frac{i^{13}}{2^{13}\pi^{13}}\,\Omega_{[i_1}{}^{i_1}\wedge\Omega_{i_2}{}^{i_2}\wedge\Omega_{i_3}{}^{i_3}\wedge\Omega_{i_4}{}^{i_4}\wedge\Omega_{i_5}{}^{i_5}\wedge\Omega_{i_6}{}^{i_6}\wedge\Omega_{i_7}{}^{i_7}\wedge\Omega_{i_8}{}^{i_8}\wedge\Omega_{i_9}{}^{i_9}\wedge\Omega_{i_{10}}{}^{i_{10}}\wedge\Omega_{i_{11}}{}^{i_{11}}\wedge\Omega_{i_{12}}{}^{i_{12}}\wedge\Omega_{i_{13}]}{}^{i_{13}} \qquad (16)\\
&= \frac{i^{13}}{2^{13}\pi^{13}13!}(+\,479{,}001{,}600\ \mathrm{tr}(\Omega^{13}) - 518{,}918{,}400\ \mathrm{tr}(\Omega^{12})\wedge\mathrm{tr}(\Omega) - 283{,}046{,}400\ \mathrm{tr}(\Omega^{11})\wedge\mathrm{tr}(\Omega^2) + 283{,}046{,}400\ \mathrm{tr}(\Omega^{11})\wedge\mathrm{tr}^2(\Omega) -\\
&\quad - 207{,}567{,}360\ \mathrm{tr}(\Omega^{10})\wedge\mathrm{tr}(\Omega^3) + 311{,}351{,}040\ \mathrm{tr}(\Omega^{10})\wedge\mathrm{tr}(\Omega^2)\wedge\mathrm{tr}(\Omega) - 103{,}783{,}680\ \mathrm{tr}(\Omega^{10})\wedge\mathrm{tr}^3(\Omega) -\\
&\quad - 172{,}972{,}800\ \mathrm{tr}(\Omega^9)\wedge\mathrm{tr}(\Omega^4) + 230{,}630{,}400\ \mathrm{tr}(\Omega^9)\wedge\mathrm{tr}(\Omega^3)\wedge\mathrm{tr}(\Omega) + 86{,}486{,}400\ \mathrm{tr}(\Omega^9)\wedge\mathrm{tr}^2(\Omega^2) -\\
&\quad - 172{,}972{,}800\ \mathrm{tr}(\Omega^9)\wedge\mathrm{tr}(\Omega^2)\wedge\mathrm{tr}^2(\Omega) + 28{,}828{,}800\ \mathrm{tr}(\Omega^9)\wedge\mathrm{tr}^4(\Omega) - 155{,}675{,}520\ \mathrm{tr}(\Omega^8)\wedge\mathrm{tr}(\Omega^5) +\\
&\quad + 194{,}594{,}400\ \mathrm{tr}(\Omega^8)\wedge\mathrm{tr}(\Omega^4)\wedge\mathrm{tr}(\Omega) + 129{,}729{,}600\ \mathrm{tr}(\Omega^8)\wedge\mathrm{tr}(\Omega^3)\wedge\mathrm{tr}(\Omega^2) - 129{,}729{,}600\ \mathrm{tr}(\Omega^8)\wedge\mathrm{tr}(\Omega^3)\wedge\mathrm{tr}^2(\Omega) -\\
&\quad - 97{,}297{,}200\ \mathrm{tr}(\Omega^8)\wedge\mathrm{tr}^2(\Omega^2)\wedge\mathrm{tr}(\Omega) + 64{,}864{,}800\ \mathrm{tr}(\Omega^8)\wedge\mathrm{tr}(\Omega^2)\wedge\mathrm{tr}^3(\Omega) - 6{,}486{,}480\ \mathrm{tr}(\Omega^8)\wedge\mathrm{tr}^5(\Omega) -\\
&\quad - 148{,}262{,}400\ \mathrm{tr}(\Omega^7)\wedge\mathrm{tr}(\Omega^6) + 177{,}914{,}880\ \mathrm{tr}(\Omega^7)\wedge\mathrm{tr}(\Omega^5)\wedge\mathrm{tr}(\Omega) + 111{,}196{,}800\ \mathrm{tr}(\Omega^7)\wedge\mathrm{tr}(\Omega^4)\wedge\mathrm{tr}(\Omega^2) -\\
&\quad - 111{,}196{,}800\ \mathrm{tr}(\Omega^7)\wedge\mathrm{tr}(\Omega^4)\wedge\mathrm{tr}^2(\Omega) + 49{,}420{,}800\ \mathrm{tr}(\Omega^7)\wedge\mathrm{tr}^2(\Omega^3) - 148{,}262{,}400\ \mathrm{tr}(\Omega^7)\wedge\mathrm{tr}(\Omega^3)\wedge\mathrm{tr}(\Omega^2)\wedge\mathrm{tr}(\Omega) +\\
&\quad + 49{,}420{,}800\ \mathrm{tr}(\Omega^7)\wedge\mathrm{tr}(\Omega^3)\wedge\mathrm{tr}^3(\Omega) - 18{,}532{,}800\ \mathrm{tr}(\Omega^7)\wedge\mathrm{tr}^3(\Omega^2) + 55{,}598{,}400\ \mathrm{tr}(\Omega^7)\wedge\mathrm{tr}^2(\Omega^2)\wedge\mathrm{tr}^2(\Omega) -\\
&\quad - 18{,}532{,}800\ \mathrm{tr}(\Omega^7)\wedge\mathrm{tr}(\Omega^2)\wedge\mathrm{tr}^4(\Omega) + 1{,}235{,}520\ \mathrm{tr}(\Omega^7)\wedge\mathrm{tr}^6(\Omega) + 86{,}486{,}400\ \mathrm{tr}^2(\Omega^6)\wedge\mathrm{tr}(\Omega) + 103{,}783{,}680\ \mathrm{tr}(\Omega^6)\wedge\mathrm{tr}(\Omega^5)\wedge\mathrm{tr}(\Omega^2) -\\
&\quad - 103{,}783{,}680\ \mathrm{tr}(\Omega^6)\wedge\mathrm{tr}(\Omega^5)\wedge\mathrm{tr}^2(\Omega) + 86{,}486{,}400\ \mathrm{tr}(\Omega^6)\wedge\mathrm{tr}(\Omega^4)\wedge\mathrm{tr}(\Omega^3) - 129{,}729{,}600\ \mathrm{tr}(\Omega^6)\wedge\mathrm{tr}(\Omega^4)\wedge\mathrm{tr}(\Omega^2)\wedge\mathrm{tr}(\Omega) +\\
&\quad + 43{,}243{,}200\ \mathrm{tr}(\Omega^6)\wedge\mathrm{tr}(\Omega^4)\wedge\mathrm{tr}^3(\Omega) - 57{,}657{,}600\ \mathrm{tr}(\Omega^6)\wedge\mathrm{tr}^2(\Omega^3)\wedge\mathrm{tr}(\Omega) - 43{,}243{,}200\ \mathrm{tr}(\Omega^6)\wedge\mathrm{tr}(\Omega^3)\wedge\mathrm{tr}^2(\Omega^2) +\\
&\quad + 86{,}486{,}400\ \mathrm{tr}(\Omega^6)\wedge\mathrm{tr}(\Omega^3)\wedge\mathrm{tr}(\Omega^2)\wedge\mathrm{tr}^2(\Omega) - 14{,}414{,}400\ \mathrm{tr}(\Omega^6)\wedge\mathrm{tr}(\Omega^3)\wedge\mathrm{tr}^4(\Omega) + 21{,}621{,}600\ \mathrm{tr}(\Omega^6)\wedge\mathrm{tr}^3(\Omega^2)\wedge\mathrm{tr}(\Omega) -\\
&\quad - 21{,}621{,}600\ \mathrm{tr}(\Omega^6)\wedge\mathrm{tr}^2(\Omega^2)\wedge\mathrm{tr}^3(\Omega) + 4{,}324{,}320\ \mathrm{tr}(\Omega^6)\wedge\mathrm{tr}(\Omega^2)\wedge\mathrm{tr}^5(\Omega) - 205{,}920\ \mathrm{tr}(\Omega^6)\wedge\mathrm{tr}^7(\Omega) + 41{,}513{,}472\ \mathrm{tr}^2(\Omega^5)\wedge\mathrm{tr}(\Omega^3) -\\
&\quad - 62{,}270{,}208\ \mathrm{tr}^2(\Omega^5)\wedge\mathrm{tr}(\Omega^2)\wedge\mathrm{tr}(\Omega) + 20{,}756{,}736\ \mathrm{tr}^2(\Omega^5)\wedge\mathrm{tr}^3(\Omega) + 38{,}918{,}880\ \mathrm{tr}(\Omega^5)\wedge\mathrm{tr}^2(\Omega^4) -\\
&\quad - 103{,}783{,}680\ \mathrm{tr}(\Omega^5)\wedge\mathrm{tr}(\Omega^4)\wedge\mathrm{tr}(\Omega^3)\wedge\mathrm{tr}(\Omega) - 38{,}918{,}880\ \mathrm{tr}(\Omega^5)\wedge\mathrm{tr}(\Omega^4)\wedge\mathrm{tr}^2(\Omega^2) + 77{,}837{,}760\ \mathrm{tr}(\Omega^5)\wedge\mathrm{tr}(\Omega^4)\wedge\mathrm{tr}(\Omega^2)\wedge\mathrm{tr}^2(\Omega) -\\
&\quad - 12{,}972{,}960\ \mathrm{tr}(\Omega^5)\wedge\mathrm{tr}(\Omega^4)\wedge\mathrm{tr}^4(\Omega) - 34{,}594{,}560\ \mathrm{tr}(\Omega^5)\wedge\mathrm{tr}^2(\Omega^3)\wedge\mathrm{tr}(\Omega^2) + 34{,}594{,}560\ \mathrm{tr}(\Omega^5)\wedge\mathrm{tr}^2(\Omega^3)\wedge\mathrm{tr}^2(\Omega) +\\
&\quad + 51{,}891{,}840\ \mathrm{tr}(\Omega^5)\wedge\mathrm{tr}(\Omega^3)\wedge\mathrm{tr}^2(\Omega^2)\wedge\mathrm{tr}(\Omega) - 34{,}594{,}560\ \mathrm{tr}(\Omega^5)\wedge\mathrm{tr}(\Omega^3)\wedge\mathrm{tr}(\Omega^2)\wedge\mathrm{tr}^3(\Omega) + 3{,}459{,}456\ \mathrm{tr}(\Omega^5)\wedge\mathrm{tr}(\Omega^3)\wedge\mathrm{tr}^5(\Omega) +\\
&\quad + 3{,}243{,}240\ \mathrm{tr}(\Omega^5)\wedge\mathrm{tr}^4(\Omega^2) - 12{,}972{,}960\ \mathrm{tr}(\Omega^5)\wedge\mathrm{tr}^3(\Omega^2)\wedge\mathrm{tr}^2(\Omega) + 6{,}486{,}480\ \mathrm{tr}(\Omega^5)\wedge\mathrm{tr}^2(\Omega^2)\wedge\mathrm{tr}^4(\Omega) -\\
&\quad - 864{,}864\ \mathrm{tr}(\Omega^5)\wedge\mathrm{tr}(\Omega^2)\wedge\mathrm{tr}^6(\Omega) + 30{,}888\ \mathrm{tr}(\Omega^5)\wedge\mathrm{tr}^8(\Omega) - 16{,}216{,}200\ \mathrm{tr}^3(\Omega^4)\wedge\mathrm{tr}(\Omega) - 32{,}432{,}400\ \mathrm{tr}^2(\Omega^4)\wedge\mathrm{tr}(\Omega^3)\wedge\mathrm{tr}(\Omega^2) +\\
&\quad + 32{,}432{,}400\ \mathrm{tr}^2(\Omega^4)\wedge\mathrm{tr}(\Omega^3)\wedge\mathrm{tr}^2(\Omega) + 24{,}324{,}300\ \mathrm{tr}^2(\Omega^4)\wedge\mathrm{tr}^2(\Omega^2)\wedge\mathrm{tr}(\Omega) - 16{,}216{,}200\ \mathrm{tr}^2(\Omega^4)\wedge\mathrm{tr}(\Omega^2)\wedge\mathrm{tr}^3(\Omega) +\\
&\quad + 1{,}621{,}620\ \mathrm{tr}^2(\Omega^4)\wedge\mathrm{tr}^5(\Omega) - 9{,}609{,}600\ \mathrm{tr}(\Omega^4)\wedge\mathrm{tr}^3(\Omega^3) + 43{,}243{,}200\ \mathrm{tr}(\Omega^4)\wedge\mathrm{tr}^2(\Omega^3)\wedge\mathrm{tr}(\Omega^2)\wedge\mathrm{tr}(\Omega) -\\
&\quad - 14{,}414{,}400\ \mathrm{tr}(\Omega^4)\wedge\mathrm{tr}^2(\Omega^3)\wedge\mathrm{tr}^3(\Omega) + 10{,}810{,}800\ \mathrm{tr}(\Omega^4)\wedge\mathrm{tr}(\Omega^3)\wedge\mathrm{tr}^3(\Omega^2) - 32{,}432{,}400\ \mathrm{tr}(\Omega^4)\wedge\mathrm{tr}(\Omega^3)\wedge\mathrm{tr}^2(\Omega^2)\wedge\mathrm{tr}^2(\Omega) +\\
&\quad + 10{,}810{,}800\ \mathrm{tr}(\Omega^4)\wedge\mathrm{tr}(\Omega^3)\wedge\mathrm{tr}(\Omega^2)\wedge\mathrm{tr}^4(\Omega) - 720{,}720\ \mathrm{tr}(\Omega^4)\wedge\mathrm{tr}(\Omega^3)\wedge\mathrm{tr}^6(\Omega) - 4{,}054{,}050\ \mathrm{tr}(\Omega^4)\wedge\mathrm{tr}^4(\Omega^2)\wedge\mathrm{tr}(\Omega) +\\
&\quad + 5{,}405{,}400\ \mathrm{tr}(\Omega^4)\wedge\mathrm{tr}^3(\Omega^2)\wedge\mathrm{tr}^3(\Omega) - 1{,}621{,}620\ \mathrm{tr}(\Omega^4)\wedge\mathrm{tr}^2(\Omega^2)\wedge\mathrm{tr}^5(\Omega) + 154{,}440\ \mathrm{tr}(\Omega^4)\wedge\mathrm{tr}(\Omega^2)\wedge\mathrm{tr}^7(\Omega) - 4290\ \mathrm{tr}(\Omega^4)\wedge\mathrm{tr}^9(\Omega) +\\
&\quad + 3{,}203{,}200\ \mathrm{tr}^4(\Omega^3)\wedge\mathrm{tr}(\Omega) + 4{,}804{,}800\ \mathrm{tr}^3(\Omega^3)\wedge\mathrm{tr}^2(\Omega^2) - 9{,}609{,}600\ \mathrm{tr}^3(\Omega^3)\wedge\mathrm{tr}(\Omega^2)\wedge\mathrm{tr}^2(\Omega) + 1{,}601{,}600\ \mathrm{tr}^3(\Omega^3)\wedge\mathrm{tr}^4(\Omega) -\\
&\quad - 7{,}207{,}200\ \mathrm{tr}^2(\Omega^3)\wedge\mathrm{tr}^3(\Omega^2)\wedge\mathrm{tr}(\Omega) + 7{,}207{,}200\ \mathrm{tr}^2(\Omega^3)\wedge\mathrm{tr}^2(\Omega^2)\wedge\mathrm{tr}^3(\Omega) - 1{,}441{,}440\ \mathrm{tr}^2(\Omega^3)\wedge\mathrm{tr}(\Omega^2)\wedge\mathrm{tr}^5(\Omega) +\\
&\quad + 68{,}640\ \mathrm{tr}^2(\Omega^3)\wedge\mathrm{tr}^7(\Omega) - 540{,}540\ \mathrm{tr}(\Omega^3)\wedge\mathrm{tr}^5(\Omega^2) + 2{,}702{,}700\ \mathrm{tr}(\Omega^3)\wedge\mathrm{tr}^4(\Omega^2)\wedge\mathrm{tr}^2(\Omega) - 1{,}801{,}800\ \mathrm{tr}(\Omega^3)\wedge\mathrm{tr}^3(\Omega^2)\wedge\mathrm{tr}^4(\Omega) +\\
&\quad + 360{,}360\ \mathrm{tr}(\Omega^3)\wedge\mathrm{tr}^2(\Omega^2)\wedge\mathrm{tr}^6(\Omega) - 25{,}740\ \mathrm{tr}(\Omega^3)\wedge\mathrm{tr}(\Omega^2)\wedge\mathrm{tr}^8(\Omega) + 572\ \mathrm{tr}(\Omega^3)\wedge\mathrm{tr}^{10}(\Omega) + 135{,}135\ \mathrm{tr}^6(\Omega^2)\wedge\mathrm{tr}(\Omega) -\\
&\quad - 270{,}270\ \mathrm{tr}^5(\Omega^2)\wedge\mathrm{tr}^3(\Omega) + 135{,}135\ \mathrm{tr}^4(\Omega^2)\wedge\mathrm{tr}^5(\Omega) - 25{,}740\ \mathrm{tr}^3(\Omega^2)\wedge\mathrm{tr}^7(\Omega) + 2145\ \mathrm{tr}^2(\Omega^2)\wedge\mathrm{tr}^9(\Omega) - 78\ \mathrm{tr}(\Omega^2)\wedge\mathrm{tr}^{11}(\Omega) + \mathrm{tr}^{13}(\Omega))
\end{aligned}$$

## Concluding Remark

For a check, note that the magnitudes of the numerical factors in the preceding expressions for $c_{(p)}$ for $0 \le p \le 13$ in Eqs. (3) through (16) add up—aside from the respective overall numerical factors—to the corresponding numbers $p!$ of covariant index permutations comprehended by $c_{(p)}$ per Eq. (1), values for which numbers appear in Table 1 (see above).